\documentclass[aps,twocolumn,groupedaddress, nofootinbib]{revtex4}
\usepackage{graphicx}
\usepackage{amsmath,amssymb}

\begin{document}

\newcommand{\etal}      {{\it et~al.}}


\title{Contamination and conditioning of the prototype double spoke cryomodule for European Spallation Source}



\author{A. Miyazaki, H. Li, K. Fransson, K. Gajewski, L. Hermansson, R. Santiago Kern, R. Wedberg, and R. Ruber}
\affiliation{Uppsala University, Uppsala, Sweden}



\begin{abstract}
A superconducting Double Spoke Resonator (DSR) is the technology of choice in a low energy section of a high power proton linear accelerator.
At the FREIA laboratory in Uppsala University, we have tested two DSRs in a prototype cryomodule for the European Spallation Source (ESS) project.
It showed that the conditioning process of these cavity packages would be the key for the series production tests.
In this paper, we present the conditioning procedure that we developed, and also describe the results with a special focus on the cross-contamination observed between two high-power couplers.
This study defines a standard conditioning recipe for the series DSR production for ESS and also for future similar projects in the world.
\end{abstract}

\maketitle

\section{Introduction}
Spoke cavities made of superconducting bulk niobium~\cite{DELAYEN1989892} are a promising accelerating structure for linear proton and heavy ion accelerators~\cite{PhysRevSTAB.6.080101, SHEPARD2006205, OLRY2006201, PhysRevSTAB.16.102001, Li_2014} in the section of around half the speed of light $c$ as reviewed in Ref.~\cite{Facco_2016}.
The European Spallation Source (ESS) project has adopted 13 cryomodules, each containing 2 Double Spoke Resonators (DSRs), in its low energy section~\cite{Garoby_2017}.
This section receives protons at 90~MeV from a normal conducting drift tube section, accelerates them to 216~MeV and sends them to the superconducting elliptical cavity sections.
The ESS DSR is optimized to have the maximum transit time factor at $\beta=v/c=0.5$ to efficiently accelerate protons from $\beta=0.4$ to $\beta=0.58$, where $v$ is the speed of protons.
This will be the first DSR operated in a proton accelerator project, and thus its experimental study is of great importance for the future proton machines.
Table~\ref{tab:ESS_param} summarizes the main ESS DSR parameters relevant to this paper.

\begin{table}[h]
  \centering
  \begin{tabular}{lc} \hline
  parameter & value \\ \hline
  frequency $f_0$ (MHz) & 352.21 \\
  operating temperature (K) & 2 \\
  accelerating gradient $E_{\rm acc}$ (MV/m) & 9 \\ 
  average power consumption $P_{\rm c}$ (W) & 2.5 \\ 
  optimal $\beta$ & 0.5 \\ 
  RF pulse length (ms) & 3.2 \\ 
  bunch length (ms) & 2.86 \\ 
  repetition rate (Hz) & 14 \\
  external quality factor $Q_{\rm ext}$ & $1.8\times 10^{5}$ \\ 
  operation band-width (Hz) & 2000 \\  \hline
  \end{tabular}
  \caption{Main parameters of the prototype ESS DSR}
  \label{tab:ESS_param}
\end{table}

In 2015, we performed low-power tests on two prototype spoke cavities with helium jackets in HNOSS (Horizontal Nugget for Operation of Superconducting Systems) at the FREIA (Facility for Research Instrumentation and Accelerator development) laboratory~\cite{LI2015, Li859430, Li906129}.
After that, in 2017, one prototype DSR with a power coupler was installed in HNOSS and was powered by a tetrode power station~\cite{LI201963, Li1162798}.
In those tests, we developed and validated the infrastructure and testing protocols.
The cavities were tested individually in HNOSS;
thus in these previous tests, we did not address phenomena occurring particularly in the cryomodule where two cavities might interfere with each other by vacuum contamination.

\begin{figure}[th]
\includegraphics[width=85mm]{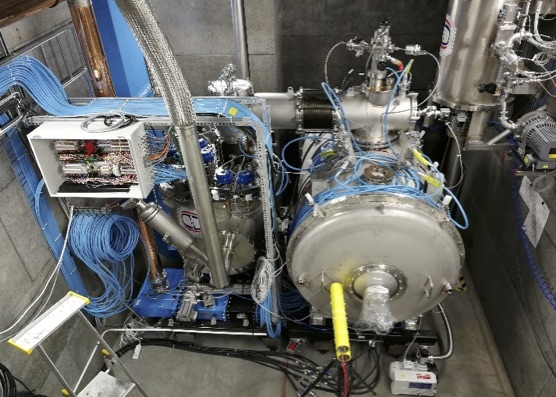}
\caption{
Photograph of the prototype cryomodule (right) and the valve box (left) in the FREIA laboratory. 
\label{fig:cm_photo}}
\end{figure}

The next milestone was to study two DSRs in a prototype cryomodule~\cite{DUTHIL2016} with more optimized parameters based on the lessons learned in 2017.
After the high power test in HNOSS, the DSR was sent to IPN Orsay, and there it was rinsed and assembled into a prototype cryomodule with another DSR.
Two power couplers were preconditioned in a dedicated conditioning bench and were assembled as well.
This prototype cryomodule was shipped to the FREIA laboratory and was tested in 2019, in the bunker as shown in Fig.~\ref{fig:cm_photo}.
The aim of the test included studying the conditioning procedure~\cite{Miyazaki:2019njx}, accelerating gradient, dynamic heat load, cold tuner, piezo actuator, diagnostics, Low-Level Radio Frequency (LLRF), and safety interlocks~\cite{Santiago-Kern1372661, Li2019}.
In this paper, we report on the new results from the conditioning process of the DSRs, which were crucial when these DSRs shared the same beam vacuum in the prototype cryomodule.

This report is organized as follows.
Section~\ref{sec:cond_proc} gives general information of the conditioning procedures performed in the FREIA laboratory.
In section~\ref{sec:coupler_cond}, we discuss the conditioning of the high power couplers.
The conditioning process and its results are followed by our findings on cross-contamination of the two couplers.
We describe this newly observed cross-contamination with promising mitigation proposals for the series production cryomodules.
Section~\ref{sec:cav_cond} is dedicated to the cavity conditioning.
Remarkably, the cross-contamination was not observed between cavities unlike in the case of the high power couplers.
In section~\ref{sec:thermal_cycle}, we describe the influence of thermal cycles on the required conditioning of the couplers and cavities.
This is of practical importance for cryomodule commissioning after installation in the accelerator.
Section~\ref{sec:conclusion} represents conclusions.

\section{General information of conditioning processes\label{sec:cond_proc}}
\begin{figure}[th]
\includegraphics[width=85mm]{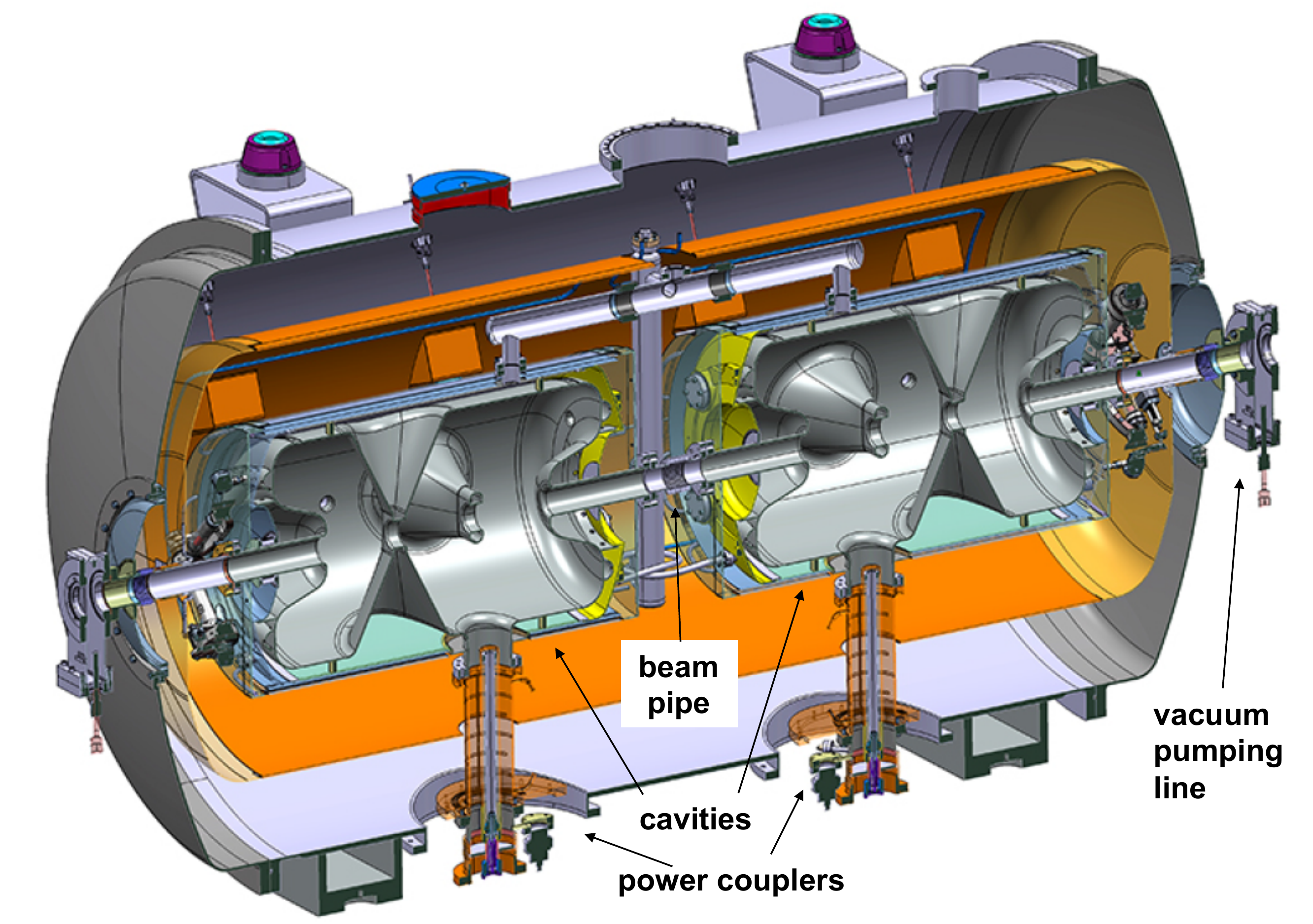}
\caption{
Schematic cross-sectional view of the DSR cryomodule~\cite{Peggs:2013sgv} (\textcopyright CNRS/IJCLab)
\label{fig:cm_sche}}
\end{figure}

Sufficient conditioning of couplers and cavities are of crucial importance in order to properly and safely evaluate and operate cavities in a cryomodule.
The dedicated tests in the FREIA laboratory are particularly useful to localize potential issues before installing the cryomodule in the accelerator tunnel, in which individual tests of each cryomodule will be practically difficult.
Figure~\ref{fig:cm_sche} shows the schematic of the ESS cryomodule for DSR, in which two cavities dressed with helium tanks are connected with a beam pipe, surrounded by a thermal shield and an isolation vacuum vessel.
High power coaxial couplers, with single TiN-coated ceramic windows, are mounted from the bottom of the cavities.
The ESS DSRs are not equipped with higher-order mode dampers.
In this prototype cryomodule testing, the beam vacuum was pumped from one side through the beam port with one turbo-molecular pump.
The major distinction from our last report on the high-power test~\cite{LI201963} was that now the two cavity packages shared the same beam vacuum.

The FREIA laboratory is equipped with two power stations dedicated to the ESS project~\cite{JOBS2017}.
During the prototype cryomodule testing, one power station was available while the other was under commissioning~\cite{Ruber1371207}.
The conditioning was performed by switching from one cavity package to the other with the available high power station, which outputs maximum 400~kW (duty ratio 4.5\%) from two combined tetrode tubes.
The RF power was transported to the cryomodule via a coaxial and rectangular waveguide system with high-power circulators and loads equipped with water cooling at a precisely controlled temperature.
Just below the cryomodule, the waveguide mode was converted to the coaxial mode by a doorknob adapter.
The stress tests of the power station and the waveguide system were included in the prototype testing program.

In the last report on one cavity testing~\cite{LI201963} in the HNOSS cryostat, we noticed a substantial dynamic heat load in the cavity compared to a vertical test result.
During the disassembly of this poorly performing cavity package, it turned out that the ceramic window was sputtered by copper.
It was thought that this could be due to contamination in the power coupler produced during its preconditioning in the dedicated bench and/or assembly onto the cavity package, which gave rise to plasma formation during the high power tests and degraded the cavity performance.
Since that test was the very first high-power experiment with the DSR cavity package, the conditioning procedure was conducted with preliminary parameters.
The issue could be mitigated by optimizing the control parameters of coupler conditioning, and
therefore, in this prototype cryomodule testing we decided to process the coupler more carefully and thoroughly.

The conditioning procedure that we developed in the FREIA laboratory was divided into coupler processing and cavity conditioning.
First, the high power couplers were conditioned at room temperature before the cooling was started.
Secondly, the couplers were reconditioned after the cavities were cooled down, in which they were tested at both the 4.2~K and 2~K temperatures.
Finally, electron activities in the cavities were conditioned at 2~K to reach the required accelerating gradient of 9~MV/m.
Every time a thermal cycle occurred, for example  for dedicated tests of the cryogenic system, 
these three steps were strictly followed to recover the functionality of the cavities.

Some instruments were used as diagnostics for the conditioning process.
One X-ray detector was placed next to the cryomodule to detect radiation generated by electron activities.
Each power coupler was equipped with a vacuum gauge next to the ceramic window.
The electron activities around the power coupler were monitored by a biased electron pick-up antenna.
Several arc detectors were also installed but were only partly used for technical reasons.
All these instruments were integrated into a control unit for monitoring and fast interlock purposes.

\section{High power coupler conditioning~\label{sec:coupler_cond}}
\subsection{Procedure of coupler conditioning}
To evaluate the cavity performance after the conditioning, we need around 120~kW forward power sent to the cavity to reach the target accelerating gradient of 9~MV/m, if the cavity resonance frequency and the generator frequency are perfectly tuned.
In the HNOSS test~\cite{LI201963} we conditioned the cavity package up to 120~kW and then measured the performance by locking the frequency with a Self-Excited-Loop (SEL).
However, in the experiment reported in this article, it was an important milestone to assess the functionality of the closed-loop feedback control of the digital LLRF system for the accelerator operation.
This required a total power around 350~kW to achieve a flat field level of 9~MV/m during the period corresponding to a beam bunch passage (2.86~ms)
\footnote{
Strictly speaking, the total power of 350~kW during the first 300~$\mu s$ is necessary to decrease the charging time of the cavity.
A similar overhead is required for the rest of the pulse for beam loading, which was reflected back to the external load during our test without proton beam.
}.
Hence, the conditioning was carried out to the maximum power of 400~kW and, therefore, the potential stress to the system, including power couplers and waveguides, was higher than the previous study.

The couplers were processed by a standing wave condition, in which avoiding to power the cavity at warm was crucial.
The power station was driven by an open loop at a generator frequency of 353~MHz, which for a cavity resonant frequency of 352~MHz, sufficiently exceeding the cavity band-width, which is around 30~kHz at warm and around 2~kHz at cold.
The cavity field was assured to be zero by monitoring the power through a pick-up antenna.

The diagnostic system is crucial to detect any suspicious activities around the coupler when it is processed by RF.
Amongst others, the vacuum gauges installed next to each power coupler's ceramic window are the most sensitive and reliable instruments for software control, and more importantly, hardware interlocks.
For the coupler conditioning, X-ray activity was not observed due to the low field level.

Table~\ref{tab:vac_param} shows the baseline parameters of the coupler conditioning defined in this prototype cryomodule testing.
The coupler started to be conditioned with a shortest pulse length of 20~$\mu$s.
At each pulse length, the power sent to the cavity package was increased from 60~dBm (1~kW) to 86~dBm (400~kW) by 0.1~dB/s.
The power was held constant if the vacuum level exceeded the lower vacuum threshold $v_{\rm L}$ to await the recovery of the vacuum level.
When substantial outgassing was detected and the vacuum level reached the higher vacuum threshold $v_{\rm H}$, the power was decreased by 3~dB until the recovery of the vacuum level to $v_{\rm L}$.
Once the power reached the maximum of 86~dBm without substantial outgassing, the pulse length was extended and power was scanned again from the minimum of 60~dBm.
Finally, the conditioning was accomplished with several power cycles at the full pulse length of 3.2~ms.
The process of controlling pulse length and power levels was automatically conducted by a LabVIEW software code.

The two vacuum threshold values $v_{\rm L}$ and $v_{\rm H}$ were manually adjusted around the typical values listed in Table~\ref{tab:vac_param} so that the conditioning period was optimized.
On top of these controlling vacuum values, a hardware interlock was defined at $v_{\rm 0}=10^{-5}$~mbar and cuts the RF immediately in case of a substantial and sudden vacuum jump.

\begin{table}[h]
  \centering
  \begin{tabular}{lc} \hline
   parameter & value \\ \hline
   minimum power (dBm) & $60$  \\
   maximum power (dBm) & $86$  \\
   pulse length ($\mu$s) & 20, 50, 100, 250, 500 \\
                         & 1000, 2000, 3200 \\ \hline
   baseline vacuum at warm (mbar) & $10^{-7}$ \\
   baseline vacuum at cold (mbar) & $10^{-9}$ \\
   hardware interlock $v_{\rm 0}$ (mbar) & $10^{-5}$ \\
   higher threshold $v_{\rm H}$ (mbar) & $\sim 5\times10^{-6}$ \\
   lower threshold $v_{\rm L}$ (mbar) & $\sim 5\times10^{-7}$ \\
   \hline
  \end{tabular}
  \caption{Baseline control parameters for the coupler conditioning}
  \label{tab:vac_param}
\end{table}

\subsection{Coupler conditioning at warm\label{sec:coupler_warm_cond}}
\begin{figure}[h]
\includegraphics[width=90mm]{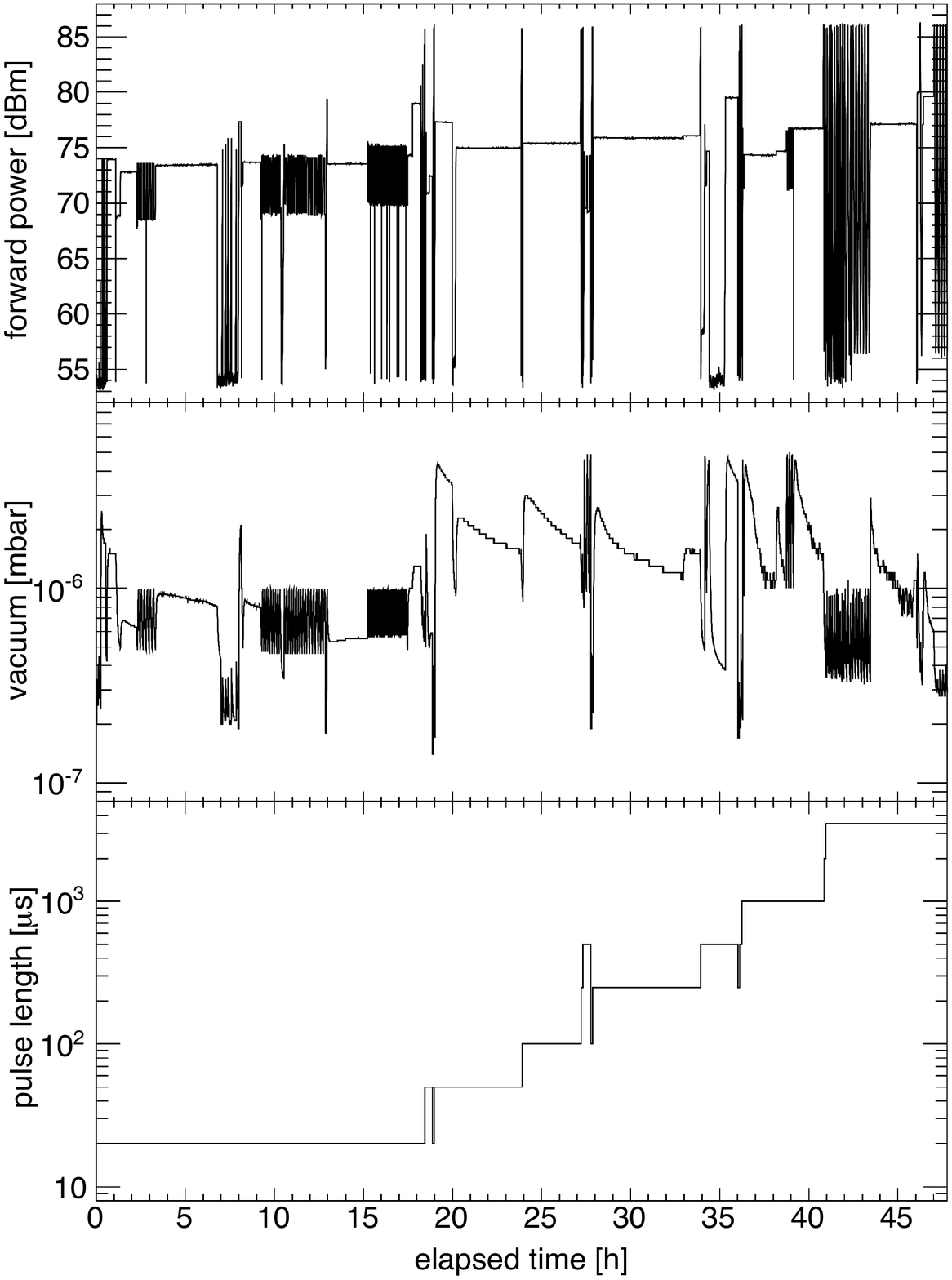}
\caption{
Evolution of coupler conditioning process at warm, with applied forward RF power controlled by the software (top), degradation of vacuum level detected by the vacuum gauge next to the coupler (middle), and pulse length of the RF (bottom). 
\label{fig:coupler_cond_warm}}
\end{figure}

Figure~\ref{fig:coupler_cond_warm} shows the evolution of one coupler conditioning at warm,
where technical interventions are disregarded from this plot.
As expected, the outgassing level rose  when either the power level or the pulse length were increased, and it slowly recovered in time.
After around 50 hours, the vacuum level was stabilized below $5\times10^{-7}$~mbar, which is comparable to the baseline vacuum at warm.
We judged that the coupler was fully conditioned at that moment.

We confirmed two power levels where the outgassing is more substantial than at other levels:
\begin{enumerate}
\item 74~dBm (25~kW) and
\item 76~dBm (40~kW) particularly with a pulse length longer than 50~$\mu$s.
\end{enumerate}
These two power levels may correspond to local multipacting barriers on the coaxial coupler and the ceramic window.
Apart from vacuum outgassing, we did not observe a clear indication of electron activities with other diagnostics, such as the electron pick-up antenna.

Although the outgassing was fully stabilized after the single coupler conditioning,
an outgassing level of one order of magnitude higher than the baseline vacuum was observed when the coupler was powered again after the conditioning of the other coupler.
This required yet another conditioning of the first coupler, which is discussed in the next section.

\subsection{Cross-contamination at warm}
During the prototype cryomodule testing, the cavity packages were conditioned one by one in a common beam vacuum.
Therefore, the multipactors removed from one cavity package could contaminate the other.
Then, during the conditioning of the other cavity package, the first cavity package could also be re-contaminated.
The reconditioning of the couplers was performed again one after another.
It is worth mentioning that this cross-contamination was only observed during the warm coupler conditioning, and correspondingly, the reconditioning was only required at warm.

\begin{figure}[h]
\includegraphics[width=90mm]{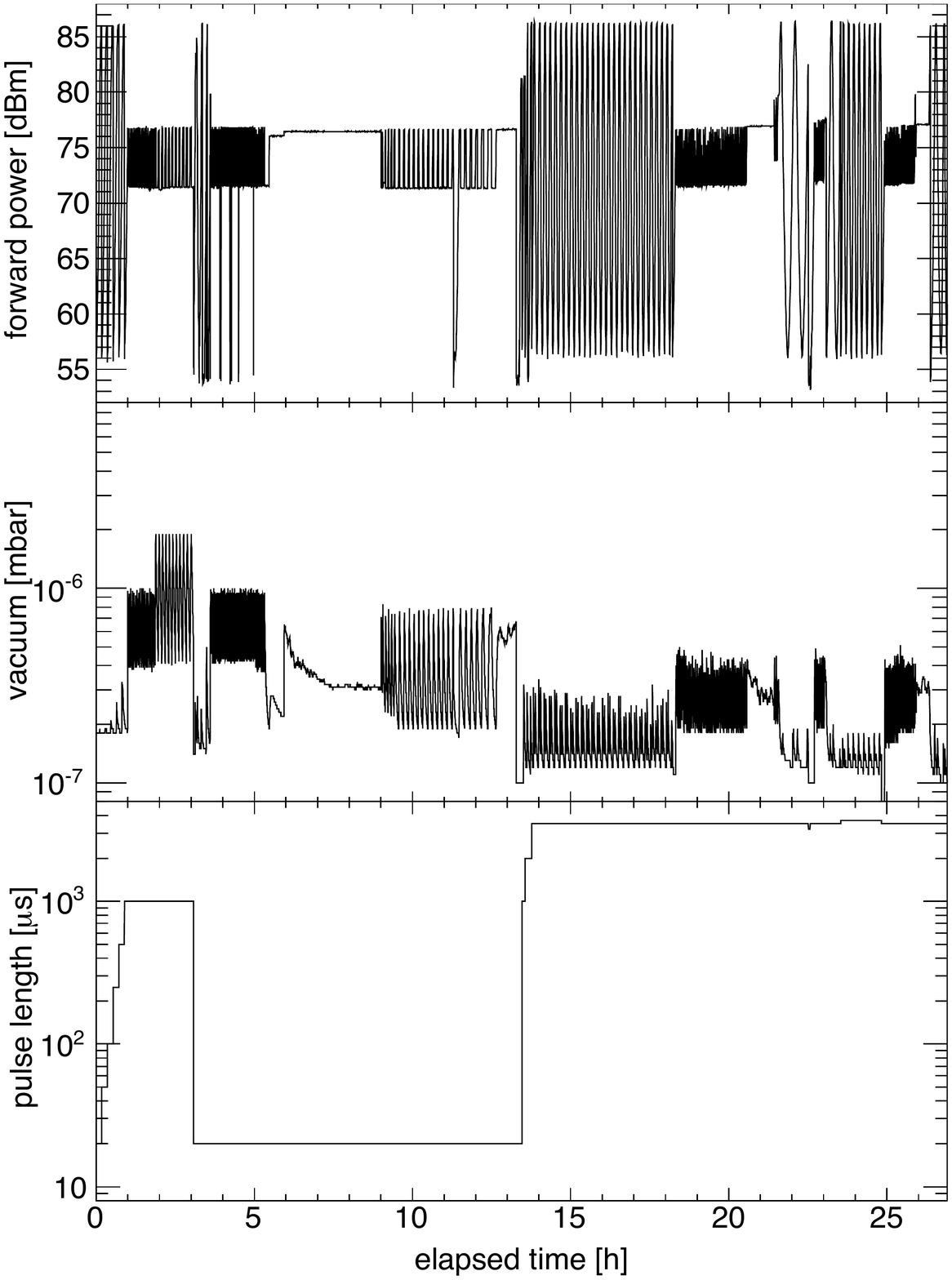}
\caption{
Evolution of coupler reconditioning process at warm, with applied forward RF power controlled by the software (top), degradation of vacuum level detected by the vacuum gauge next to the coupler (middle), and pulse length of the RF (bottom). 
\label{fig:coupler_cond_warm_2nd}}
\end{figure}

Figure~\ref{fig:coupler_cond_warm_2nd} shows the reconditioning of the coupler at warm recorded after conditioning of the other coupler.
Comparison with Fig.~\ref{fig:coupler_cond_warm} shows that the outgassing level, stabilized to below $v_{\rm L}=5\times10^{-7}$~mbar, became active again.
We confirmed this phenomenon by the vacuum rise to $v_{\rm H}$, which was set at $10^{-6}$~mbar in the software for a thorough conditioning, 
when the pulse length reached 1~ms at 2~h in Fig.~\ref{fig:coupler_cond_warm_2nd}.
At that moment, we decided to recondition more carefully to avoid further cross-contamination, and reduced the pulsed length.
This reconditioning took around 25~h, corresponding to half the initial conditioning time of 50~h.
A third conditioning was not necessary due to a rather stabilized outgassing level.
This cross-contamination happened to both couplers and therefore the total conditioning time at warm was $(50+25)\times 2\sim150$~hours.

One hypothesis of the cause of rather severe cross-contamination is the pumping path of this prototype cryomodule.
No local pumping ports were placed near the couplers, and only one turbo-molecular pump was connected through the beam port of one of the cavities with a rather long flexible tube between the cryomodule and the pump.
Depending on the conductance of the pumping port, part of the contamination can be captured by the surface of the other coupler rather than removed from the cryomodule via the vacuum pump.

\subsection{Coupler conditioning at cold}
After the coupler warm conditioning, including the reconditioning of the cross-contamination, had been accomplished, the cryomodule was cooled down.
The baseline vacuum level was around $10^{-9}$~mbar thanks to the cryogenic pumping, which also reduced the outgassing detected by vacuum gauges mounted next to the power coupler.
We tried cold conditioning at 4~K and 2~K for after a few thermal cycles.

\begin{figure}[h]
\includegraphics[width=90mm]{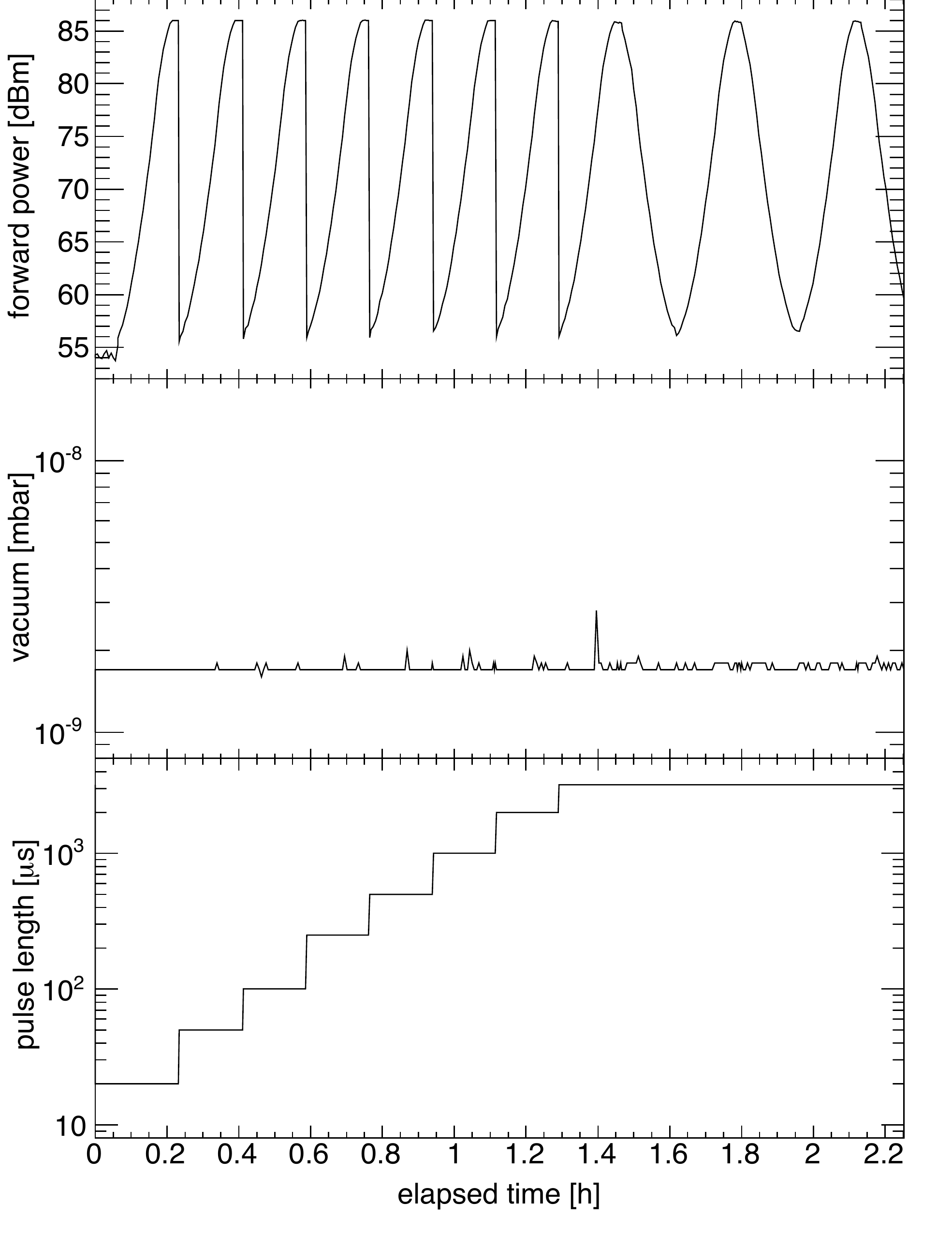}
\caption{
Evolution of coupler conditioning process at cold, with applied forward RF power controlled by the software (top), degradation of vacuum level detected by the vacuum gauge next to the coupler (middle), and pulse length of the RF (bottom). 
\label{fig:coupler_cond_cold}}
\end{figure}

Figure~\ref{fig:coupler_cond_cold} shows the typical evolution of coupler conditioning at cold.
The conditioning of one coupler at cold took only around 2 hours, independently of the ambient temperature of either 2~K or 4.2~K.
The thorough conditioning at warm mitigated the outgassing during this cold conditioning.

Unlike the conditioning at warm, the cross-contamination issue was not observed during the conditioning at cold.
This was probably because of local cryogenic pumping around the coupler.

\subsection{Conditioning period for one coupler}
Apart from the cross-contamination reconditioning at warm of 25~hours, the total time used for the coupler conditioning was $50$~(warm)~$+2$~(cold)~$=52$~h.
This 52~h can be used as a reference time of one coupler conditioning time to compare with other experiments.

In the single cavity study~\cite{LI201963}, we used 14~h for the cold conditioning while we used only 30~h for the warm conditioning;
thus, it took 44~h in total.
The warm conditioning was 12~h shorter than the test reported in this article.
In the previous test, the vacuum controlling algorithm was not identical to that used during the prototype cryomodule testing, and typical threshold parameters were one order of magnitude higher than those in Table~\ref{tab:vac_param}.
The 14~h for cold conditioning was just a default minimum period of the algorithm and was unnecessarily long.
Obviously, the procedure was not optimized for the coupler conditioning inside the cryomodule.
The warm conditioning must have been more thorough, and the cold conditioning could have been shorter.
The optimized conditioning parameters in the present report led to a promising performance after the cavity conditioning.

The required conditioning periods are the baseline of one DSR in an ESS series cryomodule.
In reality, the cross-contamination between two DSR packages takes an additional 25~h for one coupler.
It is of importance to consider a possible mitigation of the total conditioning by parallel processing.
This possibility is discussed in the next section.

\subsection{To improve the total processing time}
The experimental results so far showed that the coupler conditioning particularly at warm would be a bottleneck in the project schedule together with the cross-contamination issue.
After the prototype cryomodule test, we developed two improvements in our system to be deployed for the series cryomodule testing.
\begin{enumerate}
\item The second high power station was commissioned for simultaneous conditioning of the two cavities.
\item One additional vacuum pump will be installed at the other end of the beam port to increase the pumping capacity.
\end{enumerate}
Especially important is the second point, which will decouple the vacuum in the two cavity packages due to the low-conductance beam pipe between them.
This will help in preventing cross-contamination when outgassing occurs in one cavity package.

The simultaneous conditioning could increase twice the possible vacuum jump, which was already saturated at the controlling vacuum threshold $v_{\rm H}$.
Therefore, apart from the influence of the cross-contamination, simultaneous powering with the additional pump does not guarantee a decrease in the total conditioning process by a factor of two.
However, we can avoid switching the high power amplifier system from one cavity to the other, and the automatic software would not be interrupted.
Therefore, we will gain on the gross conditioning time by reducing such intervals.

Also, A DC biasing on the coupler antenna can potentially avoid the multipacting inside the high power coupler.
The series power coupler can be operated with this DC bias option, but its deployment in the ESS accelerator is not a baseline at the time of writing this report.
The technical design for mitigating the coupler issues observed in our study must be carefully investigated with respect to the balance between benefits and risks due to additional complexities of mechanical and operational aspects.

\section{Cavity conditioning~\label{sec:cav_cond}}
\subsection{Procedure of cavity conditioning}
After the outgassing from the couplers was stabilized at cold, the cavities were eventually powered to a certain field level.
The SEL was developed in the previous tests ~\cite{LI2015, LI201963}, with which the resonant frequency of the cavity was automatically tracked.
However, in this prototype cryomodule testing we needed to test the generator-driven operation since the SEL function will not be deployed in the ESS accelerator.
We powered the cavity by an open loop with a generator frequency adjusted to the resonant frequency of each cavity,
enabled by the wide band-width of the cavity being around 2~kHz, as shown in Table~\ref{tab:ESS_param}.

The conditioning of a cavity was carried out with short (1.0~ms) and then long (3.2~ms) RF pulses.
We started processing with the short pulse for safety and slowly increased the accelerating gradient, which was estimated by the power picked up through an antenna.
Electron activities were associated with the X-ray dose rate, and sometimes also with outgassing, and the field level was kept constant until stabilized.

The maximum field at the long pulse of 3.2~ms to be conditioned for was decided by the criteria:
\begin{enumerate}
\item up to the field at which a thermal quench happens or
\item maximum 15~MV/m if the cavity does not quench.
\end{enumerate}
Unlike the coupler conditioning, the cavity processing in the prototype cryomodule was manually performed and thoroughly checked due to a higher risk of over-powering by an unexpected cavity behavior. 
In general, the behavior of the cavities under high power RF were less predictable than for the couplers, especially considering that the DSR powering is a novel experiment in the world.

\subsection{Multipacting barriers}
The multipacting barriers have been studied during the vertical tests~\cite{IPNO_MP} and the dedicated HNOSS tests~\cite{LI201963}.
In the vertical tests, a characteristic drop of the quality factor due to the multipacting barriers were observed between 4 and 8~MV/m.
Consistently, the HNOSS test with the power coupler showed three multipacting bands: 4.5-4.8~MV/m, 5.2-5.7~MV/m, and 7.0-7.5~MV/m.

\begin{figure}[h]
\includegraphics[width=90mm]{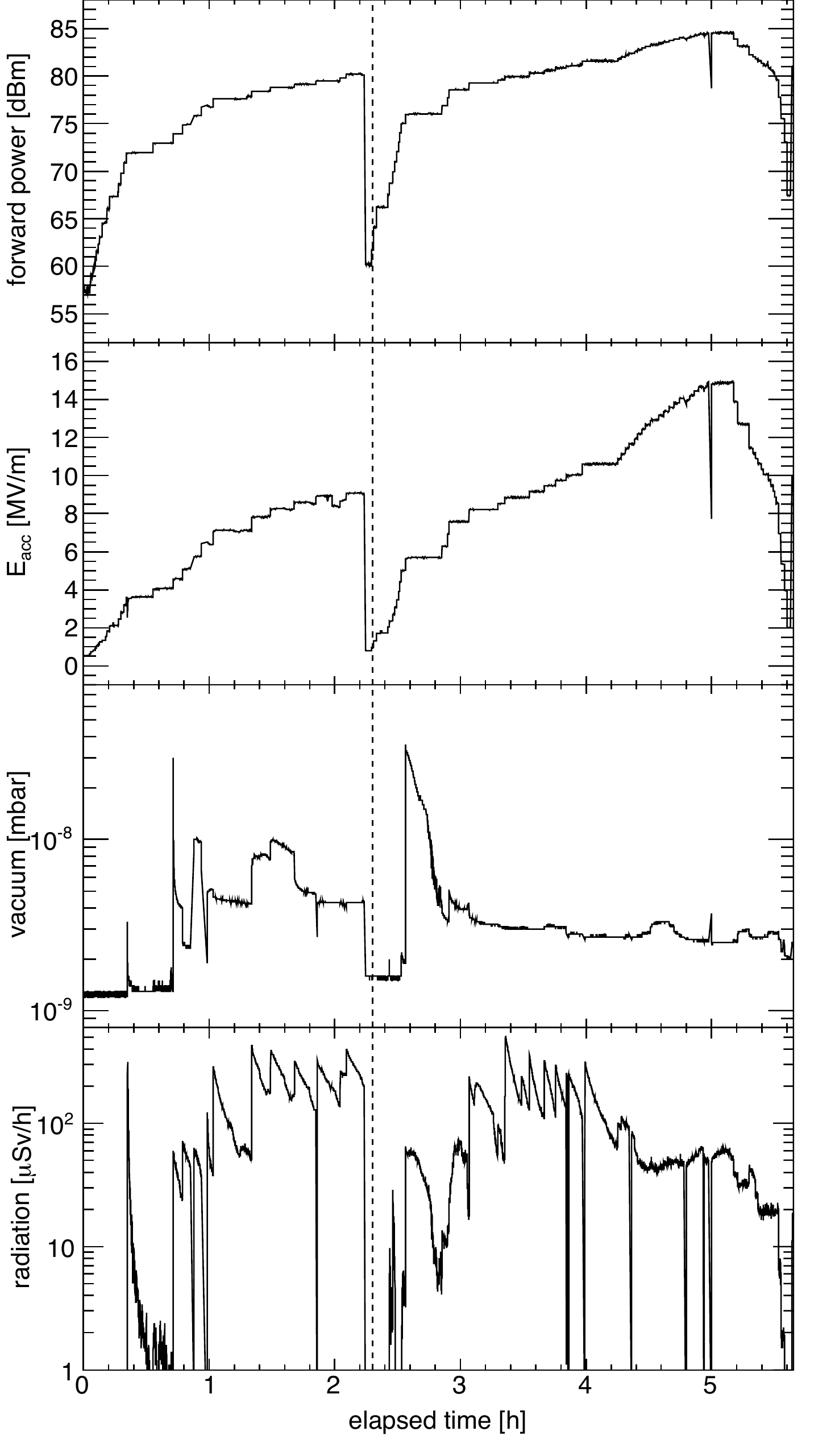}
\caption{
Evolution of cavity multipacting conditioning. From the top, forward power, accelerating gradient, vacuum level, and radiation level are shown. The left side of the vertical dashed line at 2.3~h was conditioned with pulse length of 1~ms while the right side was processed with 3.2~ms.
\label{fig:cav_MP_hist}}
\end{figure}

Figure~\ref{fig:cav_MP_hist} shows a typical result from the cavity conditioning.
Note that the small absolute scale of the radiation dose rate is not crucial for the discussion because the position of the radiation monitor was 4~m from the cavity, off-axis from the beam axis by 1.5~m, and with the other cavity acting as a radiation shield between.
The conditioning of one cavity took around 6 hours for one cavity, of which around 2 hours were used for the short pulses and the rest was with the full pulse length.

The outgassing and radiation were observed at field levels consistent with the multipacting bands, also seen in the vertical tests and the HNOSS tests.
Above 8~MV/m, one cavity (Fig.~\ref{fig:cav_MP_hist}) showed a mild field emission that was easily conditioned in short time.
The other cavity suffered from a harder field emission around similar field levels but was cleaned after a small thermal cycle up to 50~K.

Both cavities exceeded the target field level at 9~MV/m. One cavity was thermally quenched at 10.5~MV/m while the other reached 15~MV/m.
The dynamic heat load at 9~MV/m met the spec (2.5~W) in both cavities~\cite{Santiago-Kern1372661, Li2019}.

Unlike the coupler conditioning at warm, but similar to the cold coupler conditioning, no cross-contamination was observed.
This is probably because the removed gas was captured by the same cavity surface through cryogenic pumping and/or redistributed to a region geometrically more insensitive to multipacting.

\subsection{Outgassing at 74~dBm and 76~dBm}
We observed vacuum activities up to $3\times10^{-8}$~mbar at a forward power of 74~dBm with the short pulses and at 76~dBm with the long pulses (Fig.~\ref{fig:cav_MP_hist}).
The associated radiation levels were relatively low, indicating that no energetic electron activity was causing this outgassing.
These two bands were identical to the power levels where we saw outgassing during the coupler conditioning.
They were completely conditioned at warm and no activity above $3\times10^{-9}$~mbar was observed during the coupler cold conditioning (Fig.~\ref{fig:coupler_cond_cold}) before the cavity conditioning process.

A difference from the coupler conditioning was the input impedance of the cavity.
The coupler was conditioned by a standing-wave produced by a frequency mismatch of around 1~MHz above the cavity resonant frequency.
The cavity was, however,  processed by a still standing-wave in the coupler but from an over coupling condition with frequency matched within the cavity band-width.
These different impedance conditions resulted in slightly different standing-wave distributions.
These two different standing-wave patterns gave rise to an RF powering at different positions on the coupler, which were not perfectly conditioned by the coupler processing.
The outgassing was not substantial and was stabilized within 20 minutes in both cases.

\section{Influence of thermal cycles~\label{sec:thermal_cycle}}
The cryomodule was warmed up a couple of times.
It was important to check whether the contaminations were repopulated after thermal cycles
especially because of the cross-contamination mentioned before,

The small thermal cycle of the cavities from 2 K to 50 K did not influence the contamination level of the coupler at cold.
Even after a huge thermal cycle, in which the cavity temperature reached 293~K and was kept there for one month, 
the conditioning of the coupler required shorter time (a few hours) than at the first conditioning (around 77~h in total).
On the other hand, the cavity multipacting was always redistributed, and it took almost the same time (around 6~h) to finish conditioning as after the first cooling down.

\begin{figure}[h]
\includegraphics[width=90mm]{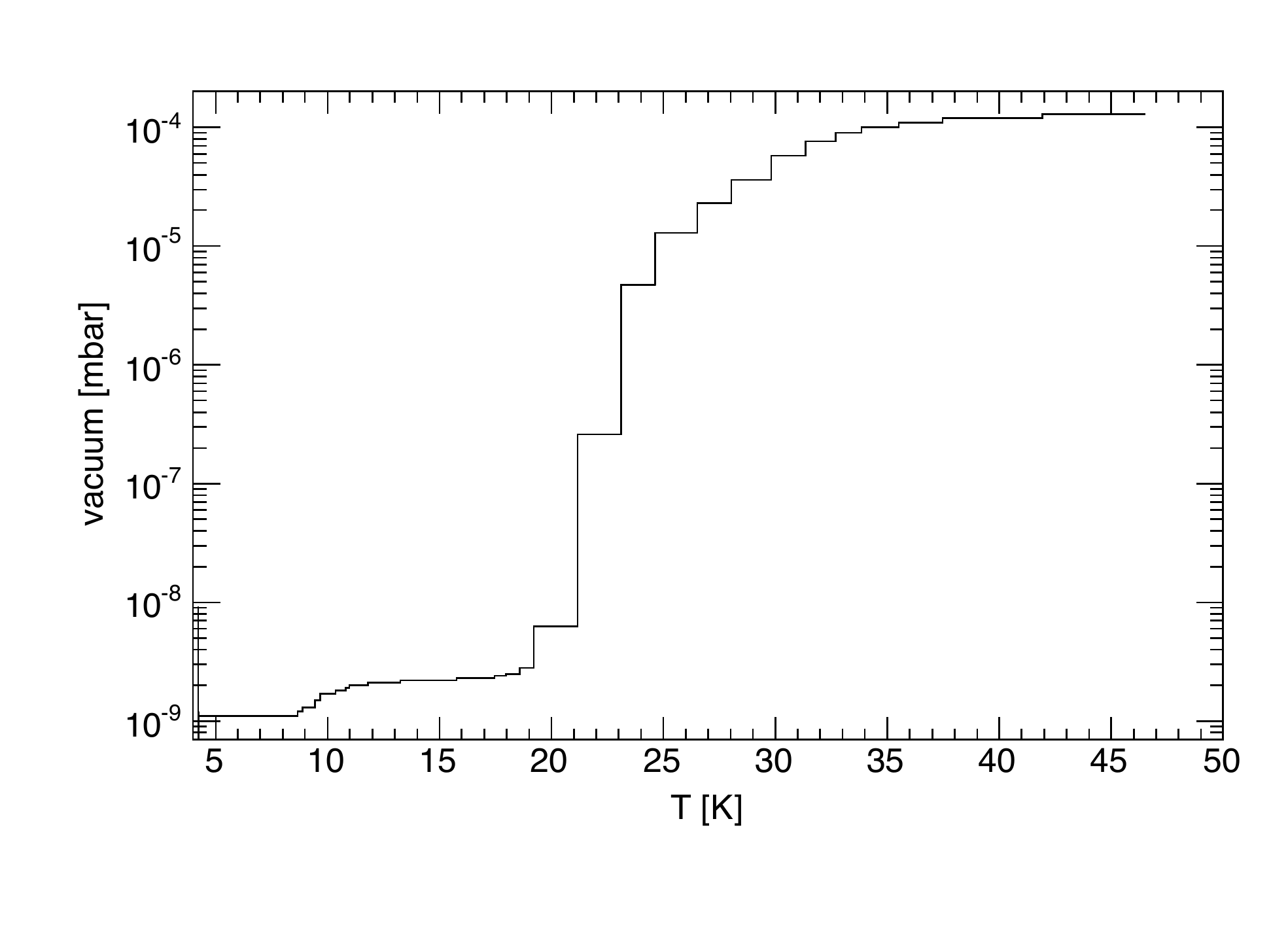}
\caption{
Outgassing during a thermal cycle when the temperature of a cavity increased from 4.2~K to room temperature.
\label{fig:thermal_cycle_outgas}}
\end{figure}

The beam vacuum was always kept below $10^{-4}$~mbar during these thermal cycles.
When the cavity package was warming up, we observed a substantial outgassing from around 20~K as shown in Fig.~\ref{fig:thermal_cycle_outgas}.
The outgassing temperature and pressure imply that nitrogen would be the dominant vapor element~\cite{Honig1960}.
A similar  outgassing always occurred in all the thermal cycles, and probably caused redistribution of the gas molecules over the cavity surface.
As the surface area of the couplers is smaller than the surface of the cavities, for capturing gas molecules, the recontamination of the couplers were not substantial compared to the repopulation of the cavity multipacting.

From these results, we can conclude that
the first coupler conditioning at warm is time consuming but the following conditionings after thermal cycles take substantially shorter time.
Although the effect of transportation has not been tested yet, we assume that the coupler conditioning time in the ESS tunnel may also be reduced thanks to the tests in the FREIA laboratory if the beam vacuum is kept under $10^{-4}$~mbar after our tests.
The cavity conditioning must be carried out every time after a thermal cycle and it takes 6~h for each cavity.
These are important feedbacks to the ESS project for optimizing conditioning procedure during RF commissioning.

\section{Conclusion~\label{sec:conclusion}}
The conditioning process of the DSRs for the ESS project was studied in a prototype cryomodule.
We developed a careful and thorough conditioning procedure to safely operate the cavities.
It was found that the coupler conditioning at warm with cross-contamination was the bottleneck in the total processing time.
We propose how to improve the warm conditioning by adding a second vacuum pump and do simultaneous conditioning of both cavity packages with an extra power station.
The cold conditioning of the coupler and the cavity did not show cross-contamination and were processed smoothly.
The multipacting barriers found for the cavities were consistent with the vertical and horizontal tests with some remarks on the possible influence from the power coupler.
After a thermal cycle both cavity and coupler conditioning must be repeated, but the coupler conditioning requires substantially shorter periods.
This study was the first experimental validation of the DSR technology and will provide a standard method for the ESS project and similar future projects.

\section*{Acknowledgements}
We would like to express our sincere gratitude to IPN Orsay and ESS for their valuable help and discussion on the prototype cryomodule testing.
We also thank all the technical staff at the FREIA laboratory.
A special thanks goes to Volker Ziemann for the useful discussion.
This project has received funding from the European Union's Horizon 2020
Research and Innovation program under Grant Agreement No 730871.

\bibliographystyle{apdrev}
\bibliography{spoke1_ver6,medium}
\end{document}